# Twisting of 2D kagomé sheets in layered intermetallics

Mekhola Sinha,[1,2] Hector K. Vivanco,[1,2] Cheng Wan,[1,2] Maxime A. Siegler,[1] Veronica J. Stewart,[1,2] Elizabeth A. Pogue,[1,2] Lucas A. Pressley,[1,2] Tanya Berry[1,2], Ziqian Wang,[3] Isaac Johnson,[3] Mingwei Chen,[3] Thao T. Tran[1,2,†], W. Adam Phelan,[1,2,‡] and Tyrel M. McQueen* [1,2,3]

[1] Department of Chemistry, The Johns Hopkins University, Baltimore, Maryland 21218, USA

[2] Institute for Quantum Matter, Department of Physics and Astronomy, The Johns Hopkins University, Baltimore, Maryland 21218, USA

[3] Department of Materials Science and Engineering, The Johns Hopkins University, Baltimore, MD 21218, USA

(*corresponding author email: *mcqueen@jhu.edu*)

**ABSTRACT:** Chemical bonding in 2D layered materials and van der Waals solids is central to understanding and harnessing their unique electronic, magnetic, optical, thermal and superconducting properties. Here we report the discovery of spontaneous, bidirectional, bilayer twisting (twist angle ~ 4.5°) in the metallic kagomé $MgCo_6Ge_6$ at T = 100(2) K via X-ray diffraction measurements, enabled by the preparation of single crystals by the Laser Bridgman method. Despite the appearance of static twisting on cooling from T ~ 300 K to 100 K, no evidence for a phase transition was found in physical properties measurements. Combined with the presence of an Einstein phonon mode contribution in the specific heat, this implies that the twisting exists at all temperatures but is thermally fluctuating at room temperature. Crystal Orbital Hamilton Population analysis demonstrates that the cooperative twisting between layers stabilizes the Co-kagomé network when coupled to strongly bonded and rigid ($Ge_2$) dimers that connect adjacent layers. Further modelling of the displacive disorder in the crystal structure shows the presence of second, Mg-deficient, stacking sequence. This alternative stacking sequence also exhibits interlayer twisting, but with a different pattern, consistent with the change in electron count due to removal of Mg. Magnetization, resistivity, and low-temperature specific heat measurements are all consistent with a Pauli paramagnetic, strongly correlated metal. Our results provide crucial insight into how chemical concepts lead to interesting electronic structures and behaviors in layered materials.

**Introduction:**

The discovery of superconductivity in an iron pnictide with a superconducting transition temperature ($T_c$) of ~ 56 K aroused tremendous interest in understanding the structure of the $ThCr_2Si_2$ structure type layered materials [1]. The compounds of the $ThCr_2Si_2$ family show high chemical flexibility to a large variety of constituents and chemical substitutions, thus numerous attempts were made to correlate the interlayer X-X (X = Si) and the tetragonal M-X (M = Cr, X =Si) bonding to the physical properties exhibited by these compounds [2-7]. This idea of applying chemical concepts to the nature of bonding in layered solids has led to the discovery of bulk Dirac cones in $Ln$AuSb ($Ln$ = La – Nd and Sm) [8] and an understanding of the rearrangement of stacking layers in van der Waals cluster magnets [9,10]. Layered intermetallic kagomés have been a perfect platform in the recent years to study the exotic phenomena associated with flat bands and Dirac-type dispersion [11-14]. Thus like the $ThCr_2Si_2$ and $Ln$AuSb materials, it is essential to explore the electronic structures of various kagomés to pave the way for experimental realization of long predicted electrical and magnetic properties of ideal 2D kagomé lattice in bulk materials.

The emergence of twist induced superconducting and insulating regions in bilayer graphene in 2018 has been a significant advancement in the emerging field of twistronics [15,16]. The stacking of two graphene sheets on top of each other and creating a small relative twist between them leads to the emergence of 'moiré bands'. These bands arise because the twist modulates the electrons tunneling in between the layers in a spatially periodic way. The moiré bands flatten at discrete set of 'magic angles' resulting in strong electronic correlations [17-20]. Since the evidence of moiré excitons have been found in van der Waals heterostructures such as $WS_2/MoSe_2$ even at bilayers with large twist angles [21,22]. Among other 2D layered systems proposed as possible candidates, theoretical predictions have shown the emergence of flat bands on interlayer twisting of 2D kagomé layers resulting in possible insulating correlated states at higher $T_c$ [23].

Here we report the discovery of spontaneous bidirectional, bilayer twisting of the Co kagomé lattice in $MgCo_6Ge_6$ of the $MT_6X_6$ (M = Mg, Li; T = Cr, Mn, Fe, Co; X = Si, Ge, Sn) family, observed by single crystal x-ray diffraction (SXRD). This twisting appears without an observable phase transition in specific heat, implying that it is present at all temperatures, but thermally fluctuating at room temperature. Integrated Crystal Orbital Hamilton Population (iCOHP) calculations have been used to identify the driving force behind this unusual behavior, which is reminiscent of the atomic reconstructions observed in twisted bilayer graphene [24].

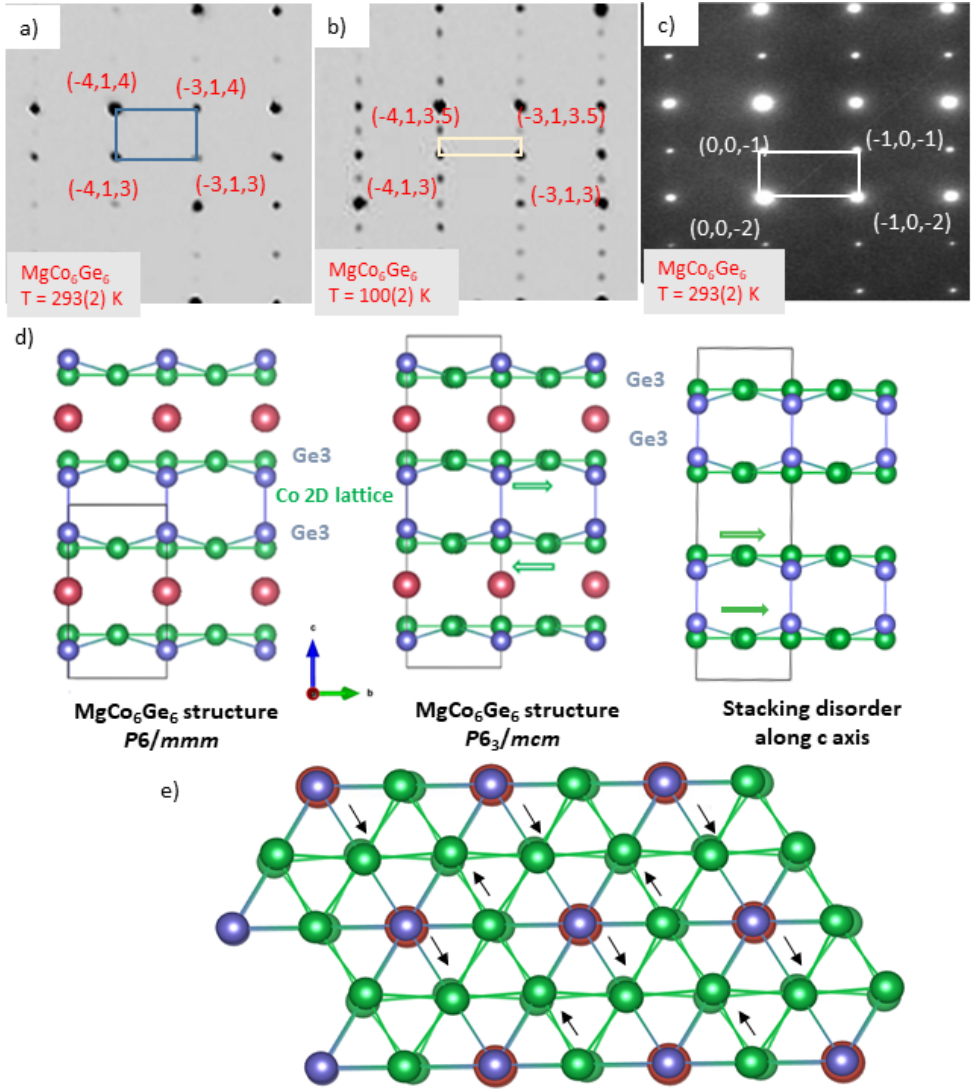

**Figure 1.** (a) SXRD precession image indexed in *P6/mmm* in the (h1l) plane for a single crystal at T = 293(2) K. The blue rectangle shows the unit cell reported for *P6/mmm* peaks which agrees with our SXRD structure solution. (b) SXRD precession images indexed in *P6$_3$/mcm* for a single crystal of MgCo$_6$Ge$_6$ at T = 100(2) K. The diffraction pattern shows doubling of the unit cell along c axis (new unit cells are shown with yellow rectangles) which can be indexed in *P6$_3$/mcm*. c) SAED image of the MgCo$_6$Ge$_6$ crystals at T = 293(2) K in the (h0l) plane. d) The bidirectional twisting of the 2D Co lattice in MgCo$_6$Ge$_6$ at T = 100(2) K, shown by green arrows. The Mg$^{2+}$ cations are showed by red spheres while the Ge atoms (except Ge3-Ge3 dumb-bells) have been omitted for clarity. The Stacking disorder observed due to the translation of the Ge3-Ge3 dumbbells accompanied by the elongation of the Ge3-Ge3 bonds in the absence of Mg$^{2+}$ cations. The structure solved in *P6/mmm* (left) does not show twisting of the Co lattice. The unit cells have been shown using rectangles. e) View of the structure in the ab plane showing the polar effect on the Co atoms due to the alternate shortening and elongation of the Co-Ge bonds in the lattice.

Magnetic and resistivity measurements done on the single crystals reveal low temperature magnetic anisotropy and Kadowaki-Woods ratio comparable to the strongly correlated metal, Na$_{0.7}$CoO$_2$ [25]. The Wilson ratio is also enhanced, and similar to strongly correlated layered cobalt oxides. Together, these results provide insight into the interplay of interlayer chemical bonding and pairwise tilting in layered materials, and how local structural relaxation can affect long-range ordering in kagomé lattices.

**Results and Discussion:**

MgCo$_6$Ge$_6$ crystals were grown by Laser Bridgman technique. Growing crystals of Mg compounds containing transition metals means dealing with the high vapor pressure of the first and the high melting temperature of the latter. Often the flux method is used to obtain single crystals. By using a flux it is possible to solve the high melting issue by obtaining a melt at lower temperatures with moderate vapor pressures that is suitable for the growth [26,27]. In some cases, the use of a solvent leads to the formation of unwanted phases; in such cases, self-flux can be useful [28]. Such fluxes can also be used as traveling solvents in a floating zone, Bridgman, or laser pedestal geometry to grow larger samples. Even using a flux or traveling solvent, growth temperatures might exceed T ~ 1200 °C which makes it harder to control the vapor pressure and almost impossible to control the decomposition of intermetallics. The vaporization can be controlled by applying inert gas pressure during the growth. Furthermore, the heating source plays an important role in the growth. The precision of laser heating enables stabilization [29] of a uniform, but small, molten volume in a Bridgman growth geometry, thus minimizing the vaporization of Mg.

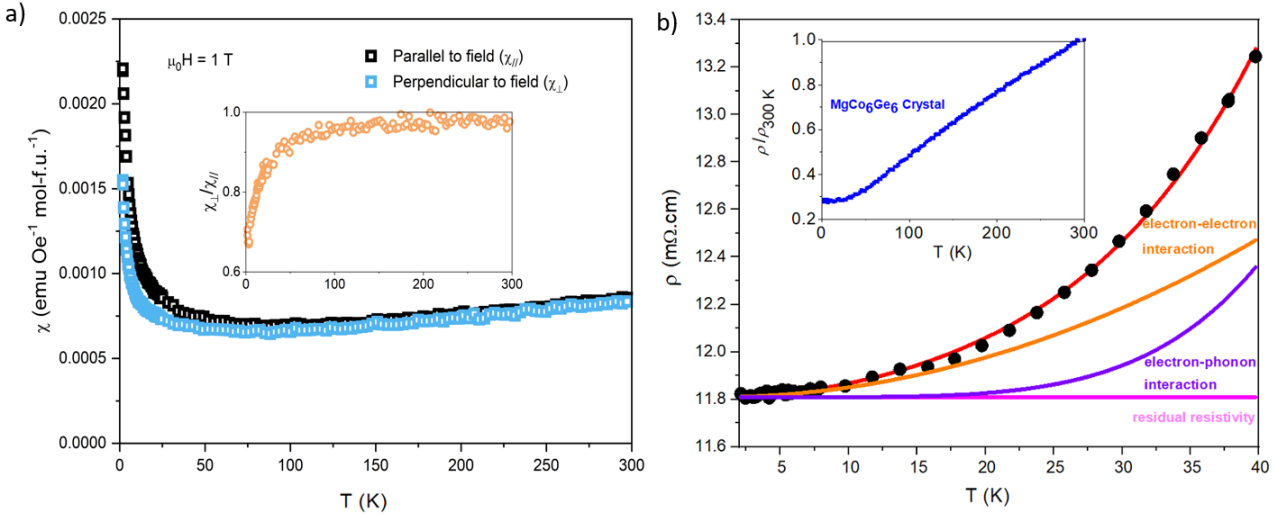

**Figure 2.** a) Magnetic susceptibility measured at $\mu_0H = 1$ T for single crystal $MgCo_6Ge_6$. The inset displays the magnetic susceptibility of the crystal [001] oriented perpendicular to the field ($\chi_\perp$) over parallel to the field ($\chi_{//}$) demonstrating the anisotropic behavior of $MgCo_6Ge_6$. The compound behaves as a paramagnet with low temperature anisotropy observed below T ~20 K. b) Resistivity measured at $\mu_0H = 0$ T for single crystal $MgCo_6Ge_6$ [001] (inset) shows no evidence of phase transition between T ~ 1.8 K and 300 K. The low temperature region from T ~1.8 to 40 K has been fit to the equation $\rho = \rho_0 + AT^2 + BT^5$ where $\rho_0 = 11.808(2)$ mΩ.cm is the residual resistivity, $A = 6(1)*10^{-5}$ mΩ.cm.K$^{-2}$ and $B = 5(3) * 10^{-10}$ mΩ.cm.K$^{-5}$ are the electron-electron and electron-phonon scattering terms respectively.

The structure of $MgCo_6Ge_6$ was characterized by single crystal X-ray diffraction, Fig 1. The room temperature (T = 293(2) K) data, Fig. 1a,d, is consistent with the previously reported crystal structure [30], and has been solved in $P6/mmm$. It consists of Co kagomé layers separated by honeycomb layers of germanium. The hexagonal holes of the Ge honeycombs are filled in alternating layers with (Ge$_2$) dimers and Mg atoms. Single crystal X-ray diffraction (SXRD) data collected at T = 100(2) K, Fig. 1b, shows doubling of the unit cell along the c axis. The low temperature structure is solved and refined in $P6_3/mcm$ (see SI, SXRD results section) Fig. 1d. We confirmed that such doubling is not present at room temperature, even locally, via selected area electron diffraction, Fig. 1c. Compared to the room temperature structure, the unit cell doubling accommodates a pronounced bidirectional twisting of the Co kagomé layers. In addition to this temperature-dependent twisting, stacking disorder (~ 5%) along the c axis accompanied by Mg$^{2+}$ vacancies coupled with elongation of Ge3-Ge3 bonds has been found in the structure at T = 100(2) K as well as at room temperature (See SI, Single crystal structure section for details).

To explore the effects of this twisting on the physical properties, multiple measurements were carried out. High temperature specific heat, discussed in greater depth below, shows no evidence for a phase transition despite this structural change, and has a low-temperature T-linear behavior expected for a metal (see SI and Fig. S1b). To assess the impact on magnetism, temperature dependent magnetic susceptibility (T ~ 1.8 to 300 K), estimated as $\chi$ = M/H with an applied field of $\mu_0H = 1$ T, of a $MgCo_6Ge_6$ crystal [001] were measured in parallel ($\chi_{//}$) and perpendicular ($\chi_\perp$) directions. A Curie-Weiss analysis (see SI) indicates the lack of local magnetism, and is consistent with the Pauli paramagnetic behavior expected from a metal, with a small contribution at the lowest temperatures from orphan/defect spins. The measurements are approximately equal at room temperature, Fig. 2a, indicative of negligible anisotropy, counter to expectations in a layered material. An anisotropy is found to develop below T ~ 20 K: $\chi_\perp$ has a different dependence with $\chi_{//}$ and $\chi_\perp$ is about 60% smaller than $\chi_{//}$ as T → 0 K (Fig. 2a inset). Isolated orphan/defect spins should be isotropic and cannot explain the anisotropic behavior. Similar anisotropic behavior has been reported in other 2D systems, such as $YCr_6Ge_6$ and $LuFe_6Ge_6$ single crystals, although its origin is not clear [31,32]. Consistent with the specific heat, no evidence of magnetic phase transition was observed down to T ~ 1.8 K.

The observed twisting is also found to minimally impact resistivity measurements, shown in Fig 2b (inset), with $MgCo_6Ge_6$ showing metallic behavior with no observable anomalies. In the same temperature region where anisotropy is found to develop in magnetization, the resistivity appears to follow a $T^2$ temperature dependence; this is indicative of the dominance of electron-electron scattering, which is quantified by the coefficient $A = 6(1)*10^{-5}$ mΩ.cm.K$^{-2}$.

Thus despite the clear change in crystal structure, heat capacity, resistivity and magnetic measurements show no indications of a phase transition between room temperature and T = 100(2) K. As such, it is likely that the bidirectional bilayer twisting is present at least up to room temperature, but with thermally activated dynamics that suppress ordering of the twists between T = 100(2) K and T = 293(2) K.

Crystal-chemical analysis gives us an insight into chemical origin of the bidirectional bilayer twisting. In the room temperature $P6/mmm$ structure, every in-plane nearest neighbor Co-Co distance is equivalent, 2.53047(8) Å. This is only slightly longer than the Co-Co distance in cobalt metal (~2.50 Å), and indicative of substantial metallic bonding within each kagomé layer. The Ge-Ge distance in each honeycomb layer (Ge1-Ge1 and Ge2-Ge2) is 2.92191(8) Å; this is substantially larger than the Ge-Ge distance in germanene (~ 2.4-2.5 Å), and indicates negligible Ge-Ge bonding within layers. Instead, the primary bonding is with cobalt: there are six Co-Ge1 bonds (2.4237(3) Å) or six Co-Ge2 bonds (2.4203(3) Å) for each honeycomb Ge atom. In contrast, the (Ge$_2$) dimers form a strong single bond, with a Ge3-Ge3 distance of 2.4997(16) Å. Each Ge3 atom completes its pseudo-tetrahedral bonding arrangement by forming bonds to adjacent Co atoms; all Co-Ge3 distances are equivalent, 2.6213(2) Å.

In the low temperature structure (solved in $P6_3/mcm$) the Co-Co, Ge-Ge, Co-Ge1 and the Co-Ge2 bond distances remain same. However, the Co-Ge3 distances show substantial changes: each (Ge$_2$) dimer has three shorter Co-Ge3 bonds (2.5091(5) Å) on one side and three longer Co-Ge3 bonds (2.7301(5) Å) on the other, forming distorted hexagonal pyramidal structure (table of contents figure). The room temperature Co-Ge3 bonds are near average of the Co-Ge3 bonds at low temperature due to the thermal activation of the bonds at higher temperature. As a normal single bond between Co-Ge in other compounds equals ~ 2.5 Å, this corresponds to strengthening three Co-Ge3 bonds and weakening three Co-Ge3 bonds. This forms an alternating pattern of long and short Co-Ge3 bonds which can be easily understood when viewed along the ab plane, Fig 1e. These effects, together, naturally explain why the Co kagomé layers on opposing sides of a (Ge$_2$) dimer twist (twist angle ~ 4.5°) in opposite directions. Further, the strong Ge3-Ge3 bonding is crucial: normally, it would not be possible to preserve local charge neutrality by lengthening or shortening all Co-Ge3 bonds to a single Ge3 atom; but as a dimer, it is possible. This also enables the kagomé layers to retain their 3-fold symmetry. The effect of the Ge3-Ge3 bonding on the interlayer twisting pattern becomes apparent when the increase in the Ge3'-Ge3' bond distance (disorder model) to 3.05(11) Å makes the Co layers twist in the same direction.

An alternative view of this structural change comes from a formal symmetry analysis: this polar disorder leads to activation of B$_1$ modes resulting in the transformation from the high temperature to the low temperature structure. Both amorphous and crystalline systems have been found to have excess low temperature specific heat than predicted by the Debye model due to low-lying optic phonons, which can arise due to proximity to bond making or breaking or directional rearrangement, such as octahedral tilting in perovskites [33], lone-pair driven disorder in pyrochlores, etc. [34]; thus a structure with evidence of temperature dependent tilt may show thermodynamic impacts of those structural components. The phonon heat capacity of a material can be modelled as a combination of Einstein and Debye modes which are optic and acoustic modes i.e. vibrations corresponding to small and large dispersions in frequency respectively. The corresponding equations are given by [34]:

$$C_{Eins} = 3sR\left(\frac{\theta_E}{T}\right)^2 \frac{\exp(\theta_E/T)}{[\exp(\theta_E/T)-1]^2}$$

$$C_{Debye} = 9sR\left(\frac{T}{\theta_D}\right)^3 \int_0^{\theta_D/T} \frac{(\theta/T)^4 \exp(\theta/T)}{[\exp(\theta/T)-1]^2} d\frac{\theta}{T}$$

where s is the oscillator strength and θ is the characteristic temperature of each of the modes. Fig. 3 shows a plot of Cp/T$^3$ vs log T; at sufficiently low temperatures Debye modes plateau at a constant value, while Einstein modes account for non-dispersing behaviour [35]. We find a definitive peak that cannot be explained by diffuse vibrational modes at T ~ 25 K for MgCo$_6$Ge$_6$. This low lying Einstein phonon mode in the heat capacity data indicates vibrationally active local modes in the structure [33].This is naturally explained as arising from thermally activated behavior of the bidirectional bilayer twisting. To see how universal this behavior is, single crystals of other materials of the MT$_6$X$_6$ family, YCr$_6$Ge$_6$ and LuFe$_6$Ge$_6$ were characterized using heat capacity measurements and X-ray diffraction. No low temperature thermal modes could be modelled in the heat capacity data explaining the lack of relaxation behavior in these materials (Fig 3, See SI heat capacity section for detailed discussion). To elucidate potential origins of this Einstein mode in specific heat, Γ point phonon energy calculations were performed on MgCo$_6$Ge$_6$ and YCr$_6$Ge$_6$. To the precision of the calculations, the distribution of modes is the same in both materials, implying that Density Functional Theory (DFT) does not adequately capture the origin of the Einstein mode. It is plausible that there is a different emergent degree of freedom in the case of MgCo$_6$Ge$_6$ that arises from the twisting of the kagomé lattice which is beyond the level DFT can capture.

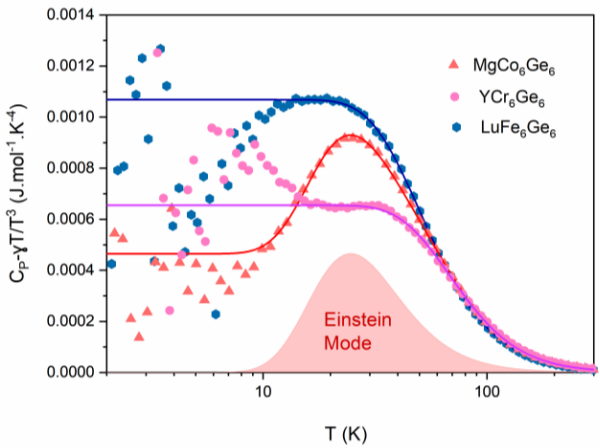

**Figure 3.** C$_P$/T$^3$ versus log of T plot after the subtraction of the electronic contribution (ɣ.T) (Fig S1b) from C$_P$ to approximate the one-dimensional phonon density of states. The solid lines show fits to the experimental data. The most noticeable difference between the fits is the contribution from the low-lying Einstein peak in the MgCo$_6$Ge$_6$ single crystal which is absent in the other two. This suggests possible presence of disorder or vibrationally active local modes in the structure.

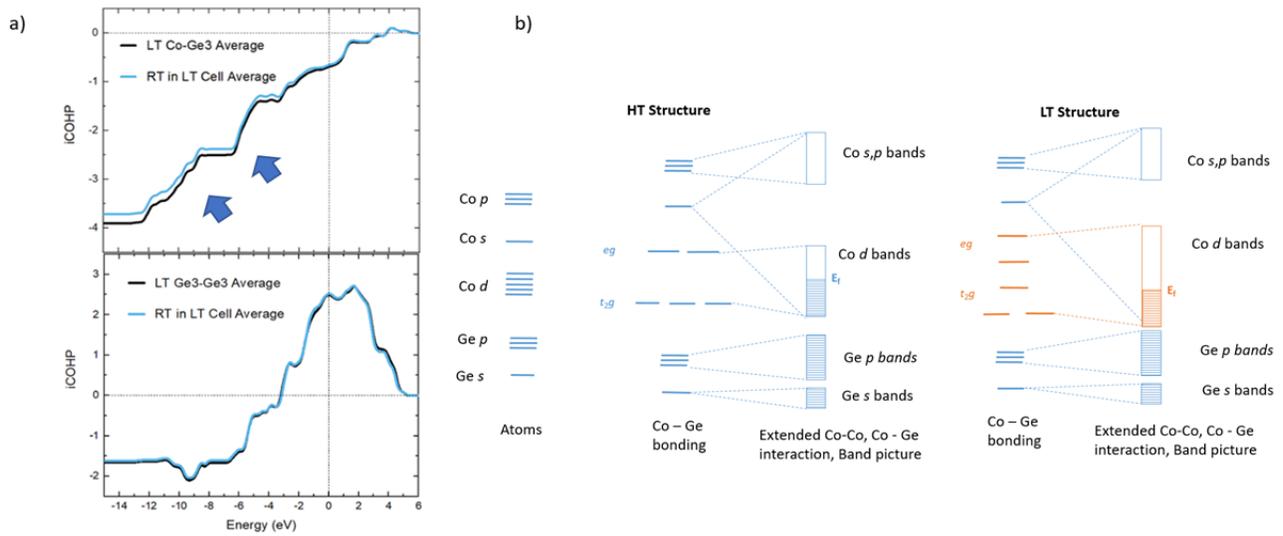

**Figure 4.** (a) Shows the iCOHP plots for both the Co-Ge3 and Ge3-Ge3 interactions, the stabilization of the Co-Ge3 interactions till Fermi level (marked by blue arrows) result in the net stabilization of the low temperature structure. The stabilization in the iCOHP interactions for Ge3-Ge3 bonds are more subtle due to the rigid nature of the dimers. The calculations for the RT structure have been done after transforming the RT structure to LT settings. (b) Qualitative band representation showing the splitting of the $d$ orbitals of Co in the low temperature structure due to the distortion in the symmetric environment which stabilizes the overall structure by decreasing the $E_f$. This agrees with the iCOHP calculations where the main stabilization factor in the structure is found to be the Co-Ge3 bonding interactions.

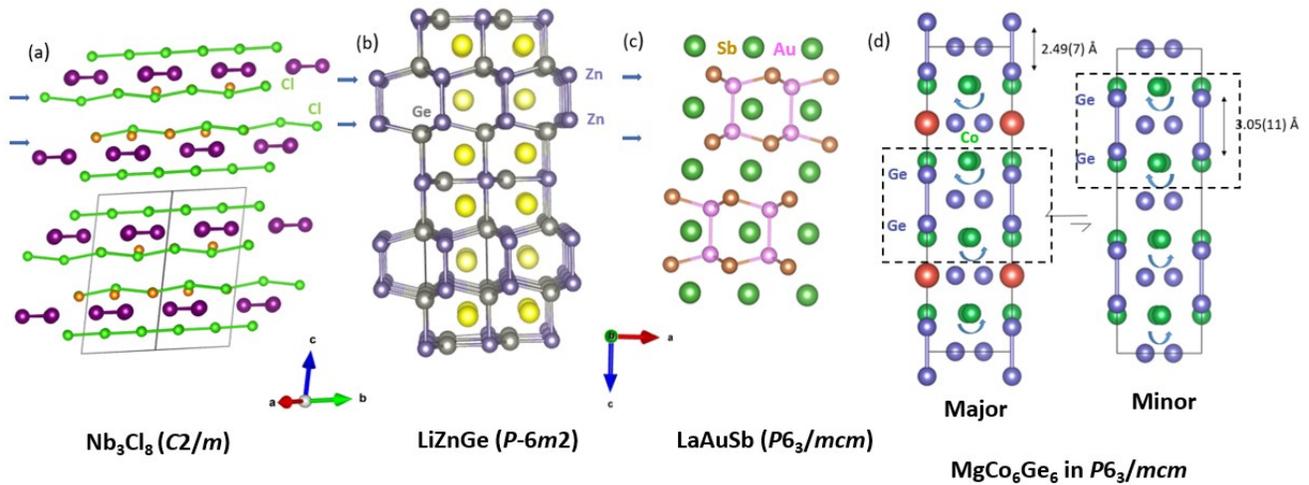

**Figure 5.** (a) The structure of Nb$_3$Cl$_8$ solved in $C2/m$ showing pairwise buckling of Cl layers (green) with blue arrows, the orange Cl atoms highlight puckering of Cl atom layers which drive closest packing and stacking order rearrangement. The Nb atoms are shown in dark purple. (b) The LiZnGe polymorph solved in $P$-$6m2$ showing paired buckling of ZnGe layers (Zn: light purple spheres are Ge: grey spheres) driven by interlayer Zn-Ge and Ge-Ge bondings. The Li atoms are shown as yellow spheres. (c) LaAuSb structure showing the interlayer Au-Au (pink) bonding which leads to the buckling of the Au-Sb (dark yellow) lattice while the green La spheres occupy the interstitial sites. (d) Left: The structure of MgCo$_6$Ge$_6$ solved in $P6_3/mcm$ along c axis. Right: Shows the minor component of the disordered model with new atomic coordinates for the Ge atoms along c axis. The Co kagomé layers rotate in the opposite directions on both sides of the Ge3-Ge3 dumbbells in the structure without disorder while opposite behaviour is observed in the minor component (highlighted with the box). The blue arrows show the pairwise buckling. The atoms contributing to the interlayer bonding and the pairwise puckering have been labelled in the figure.

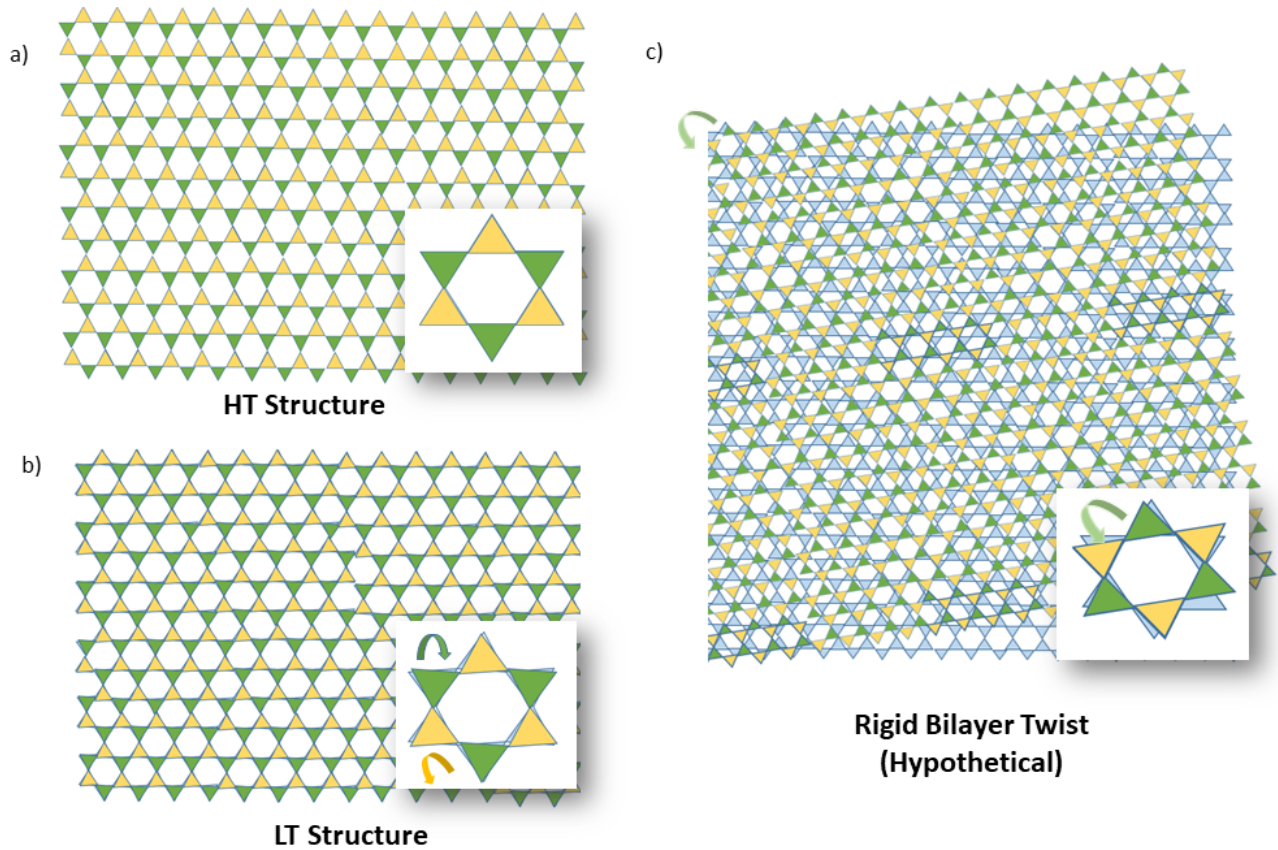

**Figure 6.** (a) The high temperature (HT) Co kagomé lattice without any tilt has eclipsed kagomé layers. (b) in the low temperature structure there are bidirectional twists of individual triangles within each layer; the minority disorder fraction exhibits tilts in the same direction. (c) If two rigid layers of the Co kagomé lattice were twisted by ~ 9 ° with respect to each other, the result would be a moiré superlattice; however, the layers are not rigid and no global rotation of one layer with respect to the next emerges, despite its presence locally.

To understand the microscopic origin of the twisting, we utilized a Crystal Orbital Hamilton Population (COHP) analysis. Integrated versus energy (iCOHP), the result is a number proportional to the energetic stabilization provided by given bonding pairs. As expected from the structural similarities, the contributions to total energy from Co-Ge1, Co-Ge2, and Mg-Ge bonds are nearly identical up to $E_f$ in both the untwisted and twisted structures. Further, consistent with the strong molecular nature of the (Ge$_2$) dimers, the Ge3-Ge3 iCOHP is also mostly unperturbed by the bilayer twist although providing stabilization at $E_f$ resulting in the net stabilization due to the twist. However, changes are observed in case of Co-Ge3: in the twisted structure, Co-Ge3 provides additional stabilization energy at all energies below $E_f$. In Fig 4a the iCOHP plot shows that the Ge3-Ge3 bonding is slightly stabilized in the low temperature structure (bond energy in the high temperature mode is 2.51595 eV while it is 2.48367 eV in the low temperature structure) while the stabilization energy for Co-Ge3 changes from -0.65161 eV to -0.6906 eV with significant stabilization of the low temperature twisted structure below $E_f$. This shows the layered structure of the Co kagomé is heavily dependent on the bonding between the Co-Ge atoms. This can be explained qualitatively by the electronic band picture shown in Fig 4b where distortion in the symmetric environment of the Co lattice in the low temperature structure leads to the splitting of the Co $d$ orbitals. The splitting results in the stabilization of the $E_f$ and the low temperature structure, a multi-site type of Jahn-Teller relaxation.

Interestingly, bilayer twisting dependent on the interlayer bonding has been observed in other layered materials as well. Figure 5 shows three different layered systems featuring pairwise twisting [See SI for detailed discussion]. The buckling of the Cl layers in the low temperature polymorph of Nb$_3$Cl$_8$ ($C2/m$)[9,10], the buckled ZnGe layers in LiZnGe ($P$-$6m2$) driven by interlayer Zn-Ge and Ge-Ge bondings [36] and the buckled Au-Sb layers in LaAuSb ($P6_3/mcm$) connected by interlayer Au-Au bonds [8] have been highlighted to emphasize on the universal pairwise twisting in layered compounds. Thus understanding of chemical bonding in these 2D layered systems throws new light on the stability and structure-physical properties correlations.

Another natural question is whether the bidirectional twisting observed in MgCo$_6$Ge$_6$ is related to twisting between layers in heterostructures that gives rise to moiré superlattices. In rigid structures like bilayer graphene, a small twist angle results into characteristic features of superlattice band structures (37-39). In the case of MgCo$_6$Ge$_6$, there are local twists between triangular units in an opposing, bidirectional fashion. Since there is no global net rotation between adjacent layers, no moiré lattice is formed. However, in the minority disorder fraction, rotations occur in the same direction; consequently, locally, there are emergent formations of small regions with a net ~ 4.5*2 = 9° twist analogous to that found in moiré superlattices, Fig 6. Put another way, the low temperature, commensurate structure can be understood as arising from applying a global rotation between adjacent layers of ~ 9°, followed by a local rotation of half of the triangles in each layer by ~ 4.5°, Fig. S2. Because the triangles within each layer are strongly bonded, the layers internally pucker and evade the formation of a moiré superlattice. This reveals a deep connection between local structural instabilities and the potential stability of moiré lattices in complex materials. The destruction of superlattice is in contrast to bilayer graphene where with application of a small global twist, we see localized in plane rotation (local structural relaxation) in the same direction. The structure is stabilized by reducing the size of the high energy domains through additional localized rotations centered around these domains [24]. However, unlike the twist in bilayer graphene no out of plane rotations were observed in $MgCo_6Ge_6$.

At the same time, there are still indications of strong electron correlations in $MgCo_6Ge_6$. The Kadowaki-Woods ratio, which compares the magnitude of electron-electron scattering in resistivity to the magnitude of the T-linear electronic contribution to the specific heat, takes on a universal value in non-correlated metals, and increases as the strength of correlations increase. We find $MgCo_6Ge_6$ has a Kadowaki-Woods ratio ($A/\gamma^2$) of 0.13 (2)*$10^{-2}$ $\mu\Omega.cm.mol^2.K^2.mJ^{-2}$, which is close to that of the strongly correlated, hexagonal layered $Na_{0.7}CoO_2$ [25, 40-41], and substantially higher than a non-correlated metal. The Wilson ratio, another metric by which to judge the strength of electron correlations [calculations in SI], is ~ 7.7 – 5.2, greater than the non-correlated metal value and similar to strongly correlated layered cobalt oxide $[BiBa_{0.66}K_{0.36}O_2]CoO_2$ which may indicate some magnetic interaction relative to the conduction electrons [42].

In summary, we discovered that the $MT_6X_6$ kagomé structures exhibit spontaneous bidirectional twisting upon cooling to low temperatures in single crystals of $MgCo_6Ge_6$. Large single crystals were obtained by using laser as the heating source in a modified Bridgman technique. SXRD data was modelled to reveal partial displacive disorder along c axis. Direction dependent low temperature (T ~ 20 K) magnetic susceptibility measurements on a single crystal of $MgCo_6Ge_6$ show anisotropy in the system similar to $LuFe_6Ge_6$ and $YCr_6Ge_6$. The transition metal lattice contributes to the bands at Fermi level making the rotation of the lattice crucial to the existence of flat bands and Dirac cones near Fermi level, Fig S3, S4, S5, much as in the case of low-buckled silicene [12]. These results demonstrate the importance of local chemical interactions in driving the behavior of twisted multilayer structures. It also opens up the possibility of the realization of flat bands in bulk specimens suitable for study by techniques that are hard to access in thin multilayer structures, such as neutron scattering.

**Methods:**

Stoichiometric amounts of germanium (Alfa Aesar, 99.97%), cobalt (Alfa Aesar, 99.7%) and magnesium (Alfa Aesar, 99%) were mixed in a molar ratio of 1:6:12 and heated at 120 °C/hr to 850 °C, held for 12 h, and then cooled to room temperature at 120 °C/hr, all in a 95% Ar, 5% $H_2$ atmosphere in a tube furnace. The sample was reground and the process was repeated three times. A zirconium foil oxygen getter was placed at the gas inlet of the tube furnace to prevent oxidation from residual $H_2O/O_2$ in the gas stream. The resulting product was purified by washing with 3 M HCl to remove residual oxide contaminants.

The synthesized powder was compacted into a graphite crucible (typically ~ 6 mm in diameter and ~ 50 mm long). A Laser Diode Floating Zone (LDFZ) furnace (Crystal Systems Inc FD-

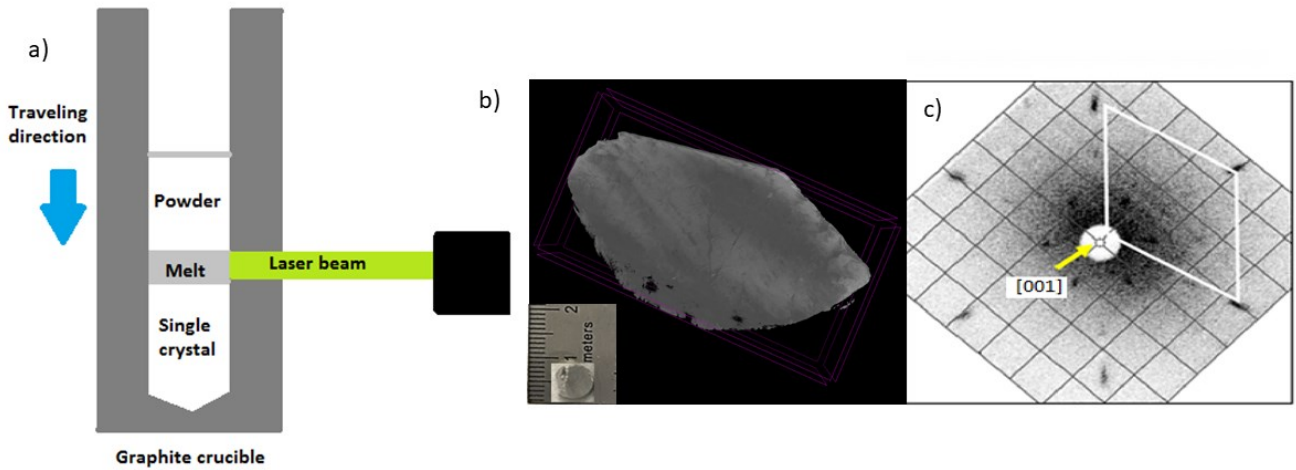

**Figure 7.** a) The experimental set up for the single crystal growth by Bridgman growth technique using GaAs lasers (only one shown for clarity) as the heating source. (b) X-ray CT scan 3D image showing no inclusions in the obtained single crystal (shown in the inset) in the micrometer range, with dark grey regions near the center being due to beam hardening artifacts. (c) Backscattered X-ray Laue image of a cut crystal taken along [001] direction showing no evidence of twinning.

FZ-5-200-VPO-PC) with 5 × 200 W GaAs lasers (976 nm) as the heating source was used for the Bridgman growth. The graphite crucible was moved downwards while the lasers were held at a fixed position (Fig 7a). Only one zone pass was required. Single crystal growth attempts under vacuum or Argon at normal pressure led to decomposition of $MgCo_6Ge_6$ into Mg and CoGe phases owing to significant vaporization of Mg at the melting temperature. The optimum growth conditions were found to be under 3 bar of dynamic argon gas pressure at ~ 40 % laser power. A high travelling rate of 10 mm/hr and narrow beam cross-section (8x4 mm) of the lasers helped to control the vaporization of Mg and yielded large (around 20 mm in length and 5 mm in diameter) air-stable single crystals of $MgCo_6Ge_6$. A rotation rate of 10 rpm was used to homogenize the melt and obtain single crystals without observable inclusions. No unexpected or unusually high safety hazards were encountered during the crystal growth process. Fig. 7b shows a 3D X-ray Micro Computed Tomography (Micro-CT) slice of an as-grown $MgCo_6Ge_6$ single crystal grown via the laser Bridgman technique. Micro-CT is a technique suitable for identification of the number, type, and distribution of micron-scale inclusions [43], with contrast provided by differing attenuation coefficients of secondary phases or crystalline boundaries. Analysis of the 3D scattering volume (SI Video 1) reveals no inclusions in the sample detected up to the limit of the camera resolution and sensitivity. There is an absence of detectable domain boundaries, consistent with a single crystal piece without twinning. Further, a backscattered X-ray Laue diffraction image collected on a piece of cut single crystal along [001] direction shows no evidence of twinning (Fig 7c).

Hexagonal plate-like single crystals of $YCr_6Ge_6$ (~ 2 mm edges) and $LuFe_6Ge_6$ (~ 1 mm edges) were obtained by previously reported flux crystal growth techniques [29, 30].

## AUTHOR INFORMATION

**Corresponding Author**
* Tyrel M. McQueen
Email: *mcqueen@jhu.edu*

**Present Addresses**
†Present Address: Department of Chemistry, Clemson University, Clemson, SC 29634, USA
‡Present Address: Los Alamos National Laboratory, New Mexico 87544, USA

**Author Contributions**
All authors have given approval to the final version of the manuscript.

**Funding Sources**
This work was supported by the David and Lucile Packard Foundation. HKV, VJS and TTT acknowledge support of the Institute for Quantum Matter, an Energy Frontier Research Center funded by the United States Department of Energy, Office of Science, Office of Basic Energy Sciences, under Award DE-SC0019331. MKS, LAP, TB and WAP acknowledge support of the Platform for the Accelerated Realization, Analysis, and Discovery of Interface Materials (DMR-1539918), a National Science Foundation Materials Innovation Platform.

## ACKNOWLEDGMENT
This work was supported by the David and Lucile Packard Foundation. HKV, VJS, EAP and TTT acknowledge support of the Institute for Quantum Matter, an Energy Frontier Research Center funded by the United States Department of Energy, Office of Science, Office of Basic Energy Sciences, under Award DE-SC0019331. MKS, LAP, TB and WAP acknowledge support of the Platform for the Accelerated Realization, Analysis, and Discovery of Interface Materials (DMR-1539918), a National Science Foundation Materials Innovation Platform. Access to the Bruker 1172 instrument was also possible via the Hopkins Extreme Materials Institute (HEMI). MKS would like to thank Dr. Chris M. Pasco for the helpful discussions regarding SXRD.

## ABBREVIATIONS
SXRD, Single Crystal X-ray Diffraction; CT, Computed Tomography; LDFZ, Laser Diode Floating Zone; DFT, Density Functional Theory; iCOHP, Integrated Crystal Orbital Hamilton Population.

## SUPPLEMENTARY INFORMATION
Experimental and characterization details, discussion on SXRD, heat capacity measurements, universal interlayer buckling and bonding in 2D systems, Wilson ratio calculation, discussion and figures from electronic band structure calculations, tables obtained from refinement to the SXRD data and fits to the heat capacity data, plots showing additional fits to the heat capacity data and figure describing local structural relaxation in the crystal structure of $MgCo_6Ge_6$.

1111